\documentclass[12pt]{article}
\usepackage{epsfig}
\usepackage{amsmath}
\usepackage{hhline}
\usepackage{amssymb}
\usepackage{times}
\usepackage{cite}

\newlength{\dinwidth}
\newlength{\dinmargin}
\begin{document}  
\newcommand{\pom}{{I\!\!P}}
\newcommand{\reg}{{I\!\!R}}
\newcommand{\tmod}{\ensuremath{|t|}}

\newcommand{\n}{1.19\pm 0.06 (stat.) \pm0.07 (syst.)}
\newcommand{\nz}{1.30\pm 0.08 (stat.)^{+0.08}_{-0.14} (syst.)}
\newcommand{\ALPHA}{1.10\pm0.03 (stat.) \pm0.04 (syst.)}
\newcommand{\ALPHAZ}{1.15\pm0.04 (stat.)^{+0.04}_{-0.07} (syst.)}

\newcommand {\gapprox}
   {\raisebox{-0.7ex}{$\stackrel {\textstyle>}{\sim}$}}
\newcommand {\lapprox}
   {\raisebox{-0.7ex}{$\stackrel {\textstyle<}{\sim}$}}
\def\gsim{\,\lower.25ex\hbox{$\scriptstyle\sim$}\kern-1.30ex%
\raise 0.55ex\hbox{$\scriptstyle >$}\,}
\def\lsim{\,\lower.25ex\hbox{$\scriptstyle\sim$}\kern-1.30ex%
\raise 0.55ex\hbox{$\scriptstyle <$}\,}
\newcommand{\pomfluxarg}{f_{\pom / p}\,(x_\pom)}
\newcommand{\dsf}{\mbox{$F_2^{D(3)}$}}
\newcommand{\dsfva}{\mbox{$F_2^{D(3)}(\beta,Q^2,x_{I\!\!P})$}}
\newcommand{\dsfvb}{\mbox{$F_2^{D(3)}(\beta,Q^2,x)$}}
\newcommand{\dsfpom}{$F_2^{I\!\!P}$}
\newcommand{\gap}{\stackrel{>}{\sim}}
\newcommand{\lap}{\stackrel{<}{\sim}}
\newcommand{\fem}{$F_2^{em}$}
\newcommand{\tsnmp}{$\tilde{\sigma}_{NC}(e^{\mp})$}
\newcommand{\tsnm}{$\tilde{\sigma}_{NC}(e^-)$}
\newcommand{\tsnp}{$\tilde{\sigma}_{NC}(e^+)$}
\newcommand{\st}{$\star$}
\newcommand{\sst}{$\star \star$}
\newcommand{\ssst}{$\star \star \star$}
\newcommand{\sssst}{$\star \star \star \star$}
\newcommand{\tw}{\theta_W}
\newcommand{\sw}{\sin{\theta_W}}
\newcommand{\cw}{\cos{\theta_W}}
\newcommand{\sww}{\sin^2{\theta_W}}
\newcommand{\cww}{\cos^2{\theta_W}}
\newcommand{\trm}{m_{\perp}}
\newcommand{\trp}{p_{\perp}}
\newcommand{\trmm}{m_{\perp}^2}
\newcommand{\trpp}{p_{\perp}^2}
\newcommand{\alp}{\alpha_s}

\newcommand{\alps}{\alpha_s}
\newcommand{\sqrts}{$\sqrt{s}$}
\newcommand{\LO}{$O(\alpha_s^0)$}
\newcommand{\Oa}{$O(\alpha_s)$}
\newcommand{\Oaa}{$O(\alpha_s^2)$}
\newcommand{\PT}{p_{\perp}}
\newcommand{\JPSI}{J/\psi}
\newcommand{\sh}{\hat{s}}
\newcommand{\uh}{\hat{u}}
\newcommand{\MP}{m_{J/\psi}}
\newcommand{\PO}{I\!\!P}
\newcommand{\xbj}{x}
\newcommand{\xpom}{\ensuremath{x_{\PO}~}}
\newcommand{\ypom}{\ensuremath{y_{\PO}}}
\newcommand{\ttbs}{\char'134}
\newcommand{\xpomlo}{3\times10^{-4}}  
\newcommand{\xpomup}{0.05}  
\newcommand{\dgr}{^\circ}
\newcommand{\pbarnt}{\,\mbox{{\rm pb$^{-1}$}}}
\newcommand{\gev}{\,\mbox{GeV}}
\newcommand{\WBoson}{\mbox{$W$}}
\newcommand{\fbarn}{\,\mbox{{\rm fb}}}
\newcommand{\fbarnt}{\,\mbox{{\rm fb$^{-1}$}}}
%
%
\newcommand{\qsq}{\ensuremath{Q^2} }
\newcommand{\gevsq}{\ensuremath{~\mathrm{GeV}^2} }
\newcommand{\et}{\ensuremath{E_t^*} }
\newcommand{\ptgamma}{\ensuremath{P_{T}^{\gamma}} }
\newcommand{\rap}{\ensuremath{\eta^*} }
\newcommand{\gp}{\ensuremath{\gamma^*}p }
\newcommand{\dsiget}{\ensuremath{{\rm d}\sigma_{ep}/{\rm d}E_t^*} }
\newcommand{\dsigrap}{\ensuremath{{\rm d}\sigma_{ep}/{\rm d}\eta^*} }

\newcommand{\alpeff}{\alpha_s^{BFKL}}
\newcommand{\wfitres}{\ensuremath{\delta=2.73 \pm 1.02~\mathrm{(stat.)}
^{+0.56}_{-0.78}~\mathrm{(syst.)}}}
\newcommand{\tfitres}{\ensuremath{n=2.60 \pm 0.19~\mathrm{(stat.)}
^{+0.03}_{-0.08}~\mathrm{(syst.)}}}
\newcommand{\alphasfitres}{\ensuremath{\alpha_S}^{Fit}=0.26 \pm 0.10
~\mathrm{(stat.)} ^{+0.05}_{-0.07}~\mathrm{(syst.)}}

\newcommand{\MYmean}{20}

\def\Journal#1#2#3#4{{#1} {\bf #2} (#3) #4}
\def\NCA{\em Nuovo Cimento}
\def\NIM{\em Nucl. Instrum. Methods}
\def\NIMA{{\em Nucl. Instrum. Methods} {\bf A}}
\def\NPB{{\em Nucl. Phys.}   {\bf B}}
\def\PLB{{\em Phys. Lett.}   {\bf B}}
\def\PRL{\em Phys. Rev. Lett.}
\def\PRD{{\em Phys. Rev.}    {\bf D}}
\def\ZPC{{\em Z. Phys.}      {\bf C}}
\def\EJC{{\em Eur. Phys. J.} {\bf C}}
\def\CPC{\em Comp. Phys. Commun.}

\begin{titlepage}

\noindent
\begin{flushleft}
{DESY 08-077    \hfill    ISSN 0418-9833} \\
{\tt December 2008}                  \\
\end{flushleft}

\vspace{2cm}


\begin{center}
\begin{Large}

{\bf Measurement of Diffractive Scattering of Photons with Large Momentum 
Transfer at HERA}

\vspace{2cm}

H1 Collaboration

\end{Large}
\end{center}

\vspace{2cm}

\begin{abstract}

The first measurement of diffractive scattering of quasi-real
photons with large momentum transfer $\gamma p \rightarrow \gamma Y$, 
where $Y$ is the proton dissociative system, is made using the H1 
detector at HERA.
The measurement is performed for initial photon virtualities $Q^2 <
0.01\gevsq$.
Single differential cross sections are measured as a function of $W$, the 
incident photon-proton centre of mass energy, 
and $t$, the square of the four-momentum
transferred at the proton vertex, in the range $175 < W < 247~\mathrm{GeV}$
and $4<\tmod <36\gevsq$. 
The $W$ dependence is well described by a model based on 
perturbative QCD using a leading logarithmic approximation 
of the BFKL evolution.
The measured $\tmod$ dependence is harder than that predicted by the model
and those observed in exclusive vector meson production.
\end{abstract}

\vspace{1.5cm}

\begin{center}
Accepted by Physics Letters B
\end{center}

\end{titlepage}

%
%
%
\begin{flushleft}

F.D.~Aaron$^{5,49}$,           
C.~Alexa$^{5}$,                
V.~Andreev$^{25}$,             
B.~Antunovic$^{11}$,           
S.~Aplin$^{11}$,               
A.~Asmone$^{33}$,              
A.~Astvatsatourov$^{4}$,       
A.~Bacchetta$^{11}$,           
S.~Backovic$^{30}$,            
A.~Baghdasaryan$^{38}$
% ...

\bigskip{\it
% ...
$ ^{4}$ Inter-University Institute for High Energies ULB-VUB, Brussels;
          Universiteit Antwerpen, Antwerpen; Belgium$^{ c}$ \\
 $ ^{5}$ National Institute for Physics and Nuclear Engineering (NIPNE) ,
          Bucharest, Romania \\
% ...
$ ^{11}$ DESY, Hamburg, Germany \\
% ...
$ ^{25}$ Lebedev Physical Institute, Moscow, Russia$^{ e}$ \\
% ...
$ ^{30}$ Faculty of Science, University of Montenegro,
          Podgorica, Montenegro$^{ e}$ \\
% ...
$ ^{33}$ Dipartimento di Fisica Universit\`a di Roma Tre
          and INFN Roma~3, Roma, Italy \\
% ...
$ ^{38}$ Yerevan Physics Institute, Yerevan, Armenia \\
% ...
$ ^{49}$ Also at Faculty of Physics, University of Bucharest,
          Bucharest, Romania \\
% ...
$ ^c$ Supported by FNRS-FWO-Vlaanderen, IISN-IIKW and IWT
      and  by Interuniversity
Attraction Poles Programme,
      Belgian Science Policy \\
% ...
$ ^e$ Supported by the Deutsche Forschungsgemeinschaft \\
% ...
}
\end{flushleft}

\newpage


%
\section{Introduction} 
\noindent

\begin{figure}[htbp]
 \begin{picture}(120,50)
  \put(1,-2){\epsfig{file=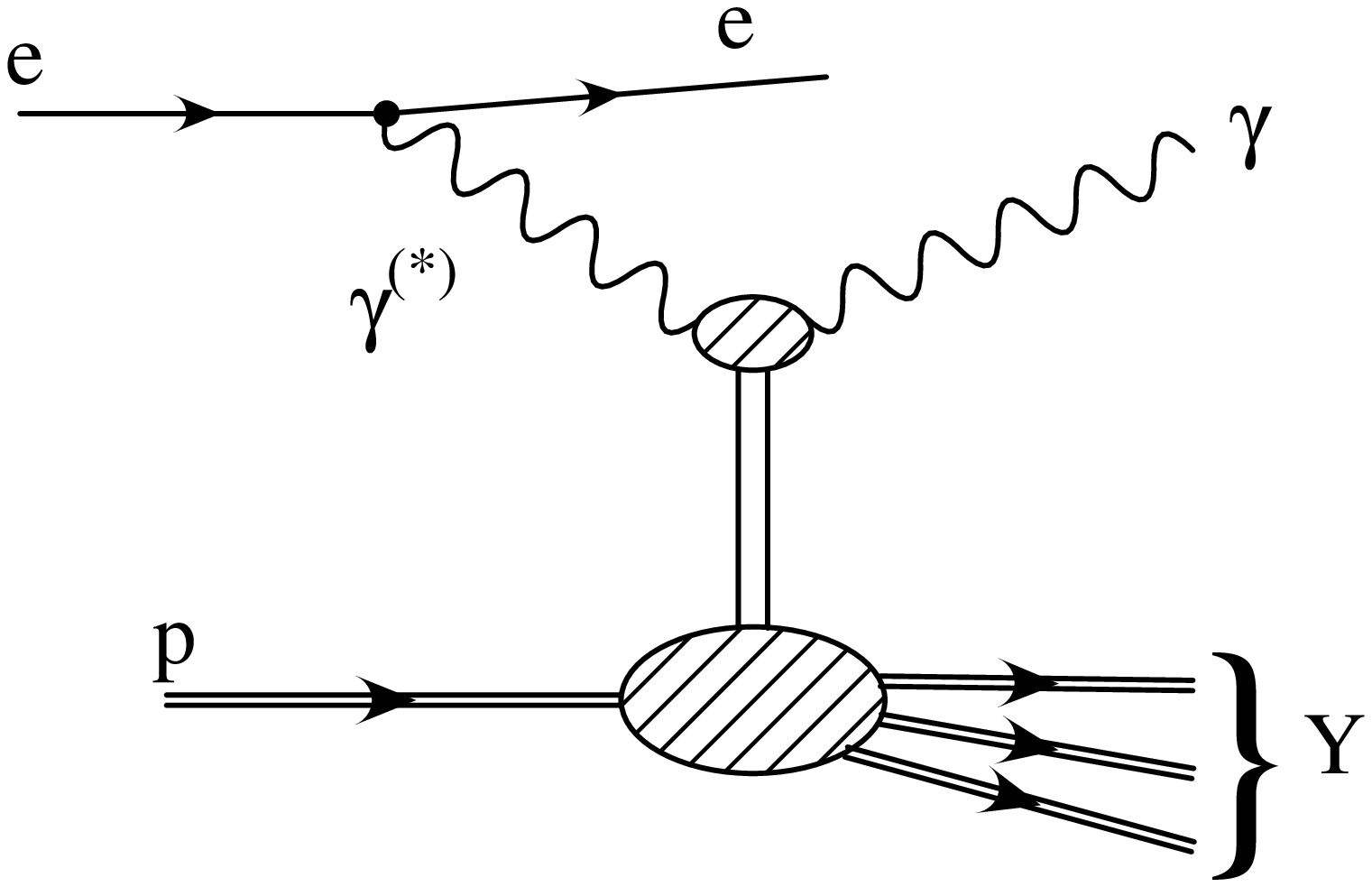,width=0.45\textwidth}}
  \put(27,20){$\xpom, t$}
  \put(4.5,44.0){$(k)$}
  \put(41.0,46.0){$(k')$}
  \put(67.0,40.5){$(p_{\gamma})$}
  \put(24.5,30.5){$(q)$}
  \put(12.0,14.0){$(p)$}
  \put(71.5,8.5){$(p_Y)$}
  \put(66.0,25.0){$\Delta \eta$}
  \put(61.0,26.5){\qbezier(0,10.0)(5,0)(0,-10.0)}
  \put(90,0){\epsfig{file=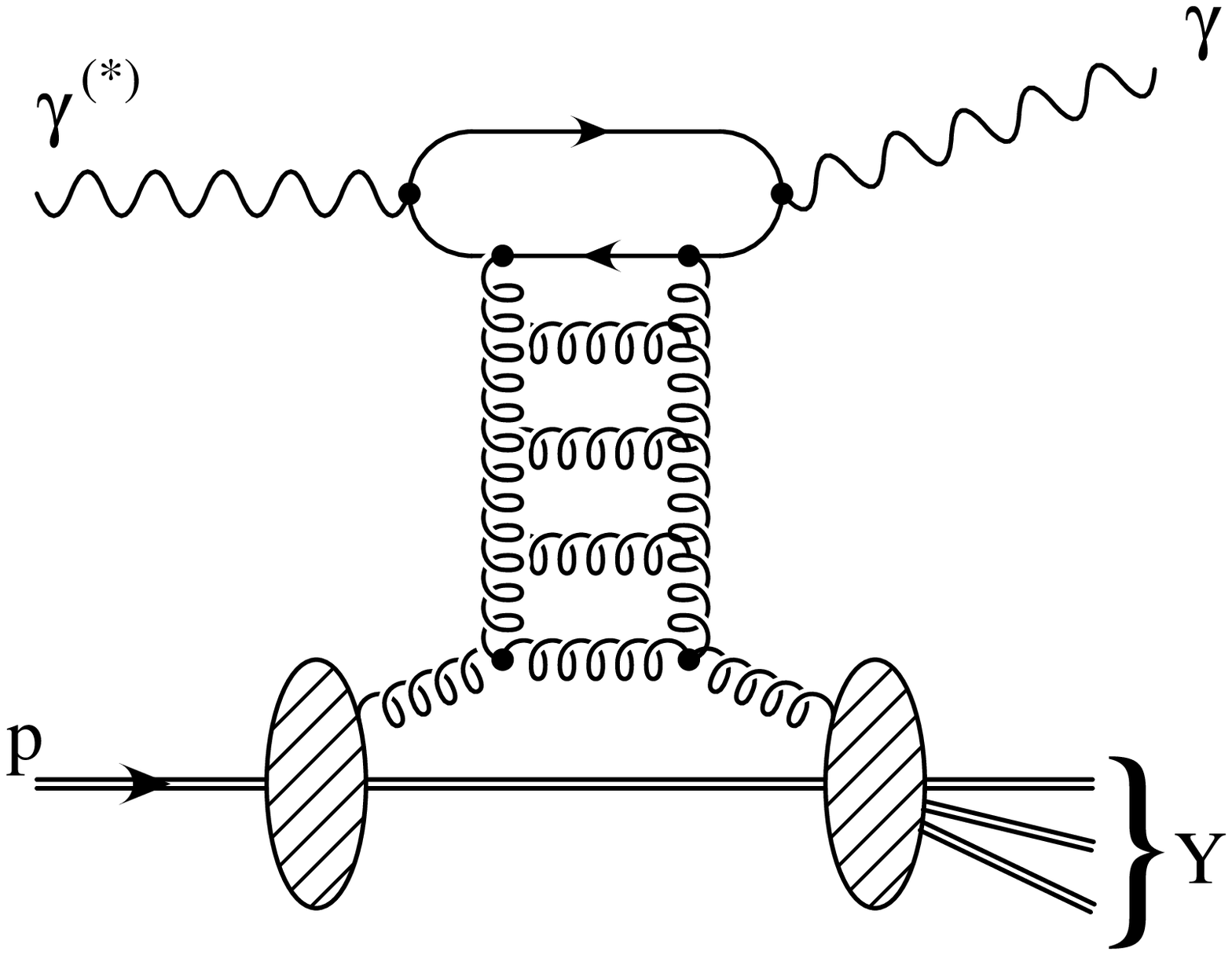,width=0.40\textwidth}}
  \put(35,-2){$a)$}
  \put(120,-2){$b)$}
 \end{picture}
 \caption{a) Schematic illustration of the $e p \rightarrow e \gamma Y$ process.
  b) Illustration of the $\gamma^{(*)} p \rightarrow \gamma Y$ process in a 
  LLA BFKL approach.
  }
 \label{fig:PhotonDiag}
\end{figure}

The study at the $ep$ collider HERA of exclusive diffractive processes 
in the presence of
a hard scale has provided insight into the parton dynamics of
the diffractive exchange. The four-momentum squared transferred at the
proton vertex, $t$, provides a relevant scale to
investigate the application of perturbative Quantum Chromodynamics
(pQCD) for $\tmod \gg \Lambda^2_{\rm QCD}$ \cite{Forshaw:1997wn}.
In this Letter, the first measurement at large $t$ ($\tmod>4$~GeV$^2$) 
of diffractive photon-proton scattering, $\gamma p \to \gamma Y$, 
where $Y$ is the proton dissociative system, is presented.
The measurement is performed at HERA by studying the reaction 
$e^+  p \to e^+  \gamma Y$ in the photoproduction regime with a large
rapidity gap between the final state photon and the proton dissociative
system~$Y$ (as illustrated in figure \ref{fig:PhotonDiag}a). 
The centre of mass energy of the system formed by 
the exchanged photon and proton is in the range $175<W<247~\mathrm{GeV}$. 
This process constitutes an extension of Deeply Virtual Compton
Scattering~\cite{:2007cz} into the large $\tmod$ and small $\qsq$ regime.

Diffractive photon scattering can be modelled in the proton rest
frame by the fluctuation of the incoming photon into a $q \bar q$ pair at 
a long distance from the proton target. 
The  $q \bar q$ pair is then involved in a hard interaction with the
proton via the exchange of a colour singlet state.
In the leading logarithmic approximation (LLA), the colour
singlet exchange is
modelled by the effective exchange of a gluon ladder (figure 
\ref{fig:PhotonDiag}b).
For sufficiently low values of Bjorken $x$ (i.e.\ large values of $W$),
the BFKL \cite{BFKL} approach is expected to
be appropriate for describing the gluon ladder. 
In the LLA BFKL model the gluon ladder couples to a single parton
(predominantly a gluon) within the proton.
The cross section depends therefore linearly on the parton distribution
in the proton.
Due to the quasi-real nature of the incoming photon ($Q^2 <
0.01\gevsq$), the transverse momentum of the final state photon,
$\ptgamma$, is entirely transferred by the gluon ladder to the parton
in the proton.
The separation in rapidity between the parton scattered by the gluon
ladder and the final state
photon is given by $\Delta \eta \simeq {\rm log} (\hat s / (\ptgamma)^2)$,
where $\hat s$ is the invariant mass of the system formed by the
incoming photon and the struck parton.
The proton remnant and the struck parton hadronise through
fragmentation processes to form the hadronic system~$Y$.
Assuming parton-hadron duality, hadrons originating from the struck parton
correspond to the particles with the largest transverse momenta
and hence are the closest in rapidity to the scattered photon.  

The present analysis complements the measurements of exclusive production of
$\rho, \phi$ and $J/\psi$ mesons at large $\tmod$
\cite{Chekanov:2002rm,Aktas:2003zi,Aktas:2006qs,Chekanov:2002xi}.
The measured $W$ and $t$ dependences of their cross sections were
found to be in agreement with LLA BFKL based calculations
\cite{Ginzburg:1985tp,Ginzburg:1996vq,Ivanov:1998jw,Evanson:1999zb,Cox:1999kv}.
For the process studied here, theoretical
calculations are simplified by the absence of a vector meson
wave function: the only non-perturbative part of the calculation is
the parton distribution functions of the proton. 
However, the cross section is suppressed relative to that of
vector meson production by the electromagnetic coupling of the
$q\bar{q}$ pair to the final state photon, making the measurement more
challenging.

In the following, measurements of the photon-proton cross sections 
are presented as a function of $W$ and differentially in $\tmod$
and are compared to predictions based on LLA BFKL 
calculations \cite{Cox:1999kv}.

\section{The H1 Detector}

A detailed description of the H1 detector can be found elsewhere
\cite{Abt:1996hi}. The following briefly describes the detector
components relevant to this analysis. A right handed coordinate system
is defined with the origin at the nominal $ep$ interaction vertex, such that
the positive $z$ axis (forward direction) corresponds to the direction
of the outgoing proton beam. The polar angle $\theta$ and transverse
momentum $P_T$ are defined with respect to the $z$ axis. The pseudorapidity is 
defined as $\eta=-\ln {\tan (\theta/2})$.

A liquid argon (LAr) calorimeter covers the polar angle range
$4^{\circ}<\theta<154^{\circ}$ with full azimuthal coverage. The
LAr calorimeter consists of both an electromagnetic section and a
hadronic section. The energy resolution for single particles measured 
in a test beam is $\sigma_E/E = 12\% / \sqrt{E/\mathrm{GeV}} \oplus 1\%$ for
electrons and $\sigma_E/E = 50\% / \sqrt{E/\mathrm{GeV}} \oplus 2\%$
for hadrons\cite{Andrieu:1994yn}. 
The polar angle region $153^{\circ} < \theta < 177^{\circ}$
is covered by a lead-scintillating fibre calorimeter, the
SpaCal \cite{Appuhn:1996na}, with both electromagnetic and hadronic sections. 
The SpaCal has an energy resolution for electromagnetic showers of
$\sigma_E/E = 7\% / \sqrt{E/\mathrm{GeV}} \oplus 1\%$.

Charged particles are detected in the polar angle ranges
$15^{\circ}<\theta<165^{\circ}$ by the Central Track Detector (CTD) and
$5^{\circ}<\theta< 25^{\circ}$ by the Forward Track Detector (FTD).
The CTD comprises two large cylindrical jet drift chambers,
providing precise track measurements in the $r-\phi$ plane, supplemented by
two $z$-chambers and two multi-wire proportional chambers 
arranged concentrically around the beam-line. 
The CTD is complemented by a silicon vertex 
detector \cite{Pitzl:2000wz}
covering the range $30^{\circ}<\theta< 150^{\circ}$.
The FTD consists of layers of planar and radial drift chambers to provide
measurement of the $\theta$ and $\phi$ angles of forward tracks,
respectively. 
The trackers and the calorimeters are operated  within a solenoidal
magnetic field of $1.16$ T.

The luminosity is determined from the rate of Bethe-Heitler events,
$ep \rightarrow ep\gamma$, measured using
a \v{C}erenkov crystal calorimeter, the Photon Detector (PD), situated 
near the HERA beam pipes at $z = -103$~m. 
A second \v{C}erenkov crystal calorimeter, the Electron Tagger (ET),
with an energy
resolution of $\sigma_E/E = 17\% / \sqrt{E/\mathrm{GeV}} \oplus 1 \%$,
is located at $z = -33$~m.
The ET measures the energy of the positron when scattered through an 
angle of less than $5$ mrad in the energy range $8 < E < 20$~GeV. 
The detection of the scattered positron in the ET ensures 
that the exchanged photon is quasi-real with $Q^2<0.01~\mathrm{GeV^2}$.

\section{Event Kinematics and Selection} 

Following the notation introduced in
figure~\ref{fig:PhotonDiag}a, the scattering process,
$e^+ p \rightarrow e^+ \gamma Y$, is described by
the usual Deep Inelastic Scattering (DIS) kinematic variables:
\begin{equation}
 \nonumber
\qsq=-q^2=-(k'-k)^2, ~~~~~\: y = \frac{p \cdot q}{p \cdot k},
\end{equation}
where $k$, $p$, $k'$ and $q$ are the four-momentum of the incoming positron, the incoming proton, the scattered positron
and the exchanged photon, respectively.
The variable $Q^2$ is the virtuality of the exchanged photon and $y$ is the
inelasticity of the $ep$ interaction, corresponding to the relative 
energy loss of the scattered
positron in the proton rest frame.
The $ep$ centre of mass energy squared is given by $s=(k+p)^2$ and
the $\gamma p$ centre of mass energy squared is $W^2=(q+p)^2 \simeq
ys$.
In addition, the longitudinal momentum
fraction of the diffractive exchange (called the Pomeron in the Regge model~\cite{Regge}) with respect to the proton is defined as:
\begin{equation}
 \nonumber
\xpom = \frac{q \cdot (p-p_Y)}{q \cdot p} \, , 
\end{equation}
where $p_Y$ is the four-momentum of the $Y$ system.
The elasticity of the $\gamma p$ interaction, which can be seen as the 
fractional energy of the
exchanged photon transferred to the final state photon, is given by
$1-\ypom$, where:
\begin{equation}
 \nonumber
\ypom = \frac{p \cdot (q-p_{\gamma})}{p \cdot q} \, ,
\end{equation}
$p_\gamma$ being the four-momentum of the final state photon.
Finally the square of the four-momentum transfer across the
diffractive exchange is given by:
\begin{equation}
 \nonumber
t=(q-p_{\gamma})^2=(p-p_Y)^2.
\end{equation}

The data used in this analysis were collected with the H1 detector during
the $1999-2000$ running period, when positrons of energy 
$E_e=27.6~\rm{GeV}$
collided with protons of energy $E_p=920~\rm{GeV}$ in the HERA accelerator. 
The data sample corresponds to an integrated luminosity of $46.2$ pb$^{-1}$. 
More details on the present analysis can be found in~\cite{Hreus08}.

The data were recorded using a combination of two triggers. 
Both triggers select an electromagnetically interacting particle
in the SpaCal which corresponds
to the scattered photon candidate. 
One of the triggers requires in addition an energy deposit 
in the ET, corresponding to the scattered positron.
The effective trigger efficiency, including time dependent downscale factors,
amounts to approximately $50\%$.

The reconstruction of the kinematic quantities used in the following are
approximations valid in the limit of small scattering angles
of the positron and small transverse momentum of the $Y$ system compared to 
its longitudinal momentum.
The quantity $y$, and hence $W$, is calculated from 
the scattered positron energy, $E_{e'}$, measured in the ET using the
relation $y=1-E_{e'}/E_e$. 
The relative resolution of $W$ is $4\%$. 
To avoid regions of low ET acceptance, the energy of the scattered positron is
limited to $11 < E_{e'}< 19$ GeV, corresponding to $175 <W< 247\ {\rm GeV}$.
In addition, to suppress backgrounds 
from processes occurring in coincidence with Bethe-Heitler events, 
it is required that no energy deposits above
the noise threshold are measured in the PD.

Photon candidates are selected from energy clusters
with small radii detected in the electromagnetic
section of the SpaCal.
If significant energy is measured behind 
the cluster in the hadronic section of the SpaCal, the event is rejected.
Events with more than one cluster above the noise level in the 
SpaCal are also rejected.
To reduce the background from charged particles, the cluster of the
photon candidate must have no associated track in the CTD.
The photon candidates are furthermore required
to have an energy $E_{\gamma}>8~\mathrm{GeV}$ and 
a transverse momentum $\ptgamma>2~\mathrm{GeV}$. 

The hadronic final state $Y$ is reconstructed using a combination of
tracking and calorimetric information. 
The difference between the total energy $E$ and the longitudinal
component of the total momentum $P_z$, as calculated from the scattered
positron, the final state photon and the hadronic system~$Y$, is
restricted to $49<\Sigma(E-P_z)<61$ GeV.
This requirement suppresses non-$ep$ induced background and 
background due to the overlap during the same bunch crossing of a
Bethe-Heitler event with a DIS event. For a fully contained $ep$ event
the relation \linebreak $\Sigma(E-P_z)= 2 E_e = 55$~GeV holds.

In the case that charged particles from the final state are detected by 
the tracking detector, allowing the primary vertex to be reconstructed, the 
$z$ coordinate of the vertex
has to satisfy $|z| < 35$~cm. For $25\%$ of the
selected events, all charged particles of the final state
lie outside the detector acceptance and no event vertex
is reconstructed. 
In this case, the time averaged vertex position is used for the kinematic 
reconstruction.

The kinematic variable $\tmod$ is reconstructed as: 
\begin{equation}
 \nonumber
 \tmod = (\ptgamma)^2 \, ,
\end{equation}
with a relative resolution of $11\%$. 
The longitudinal momentum fraction of the diffractive exchange with
respect to the proton is reconstructed using the formula:
\begin{equation}
 \nonumber
\xpom \simeq \frac{(E+P_z)_\gamma}{2E_{p}},
\end{equation}   
where $(E+P_z)_\gamma$ is the sum of the energy and longitudinal 
momentum of the final state photon.
The event inelasticity of the $\gamma p$ interaction is reconstructed as:
\begin{equation}
 \nonumber
\ypom \simeq \frac{\sum_Y(E-P_z)}{2(E_e - E_{e'})},
\end{equation}
where the summation is performed on all
detected hadronic final state particles in the event, i.e. all
measured particles except the scattered positron and the final state
photon. 
This reconstruction method has the 
advantage that the loss of particles 
along the forward beam pipe only has a marginal effect on the reconstruction of $\ypom$. 
Diffractive events are selected by requiring that $\ypom < 0.05$.
The cut value ensures a large pseudorapidity gap, $\Delta \eta$, between the
photon and the proton dissociative system~$Y$, since 
$\ypom \simeq e^{-\Delta \eta}$. 

Finally, to reduce contamination of the signal by non-diffractive background,
it is required that the difference in pseudorapidity
between the photon and the closest final state hadron satisfies 
$\Delta \eta>2$.
This rapidity gap is inferred from the absence of activity in the
relevant detector parts, i.e.\ absence of any track or a cluster of energy deposits
above the noise threshold of $400$ MeV in the LAr calorimeter.~\footnote{
The noise level of the LAr calorimeter is about $10$ MeV per cell and
the average number of cells per cluster is typically $60$.}

After applying all selection criteria, 240 events remain in the data
sample.

\section{Monte Carlo Simulations and Comparison to Data}
\label{sec:mc}

Monte Carlo (MC) simulations are used to correct the data for effects of
detector acceptance and efficiency, to estimate the
background and to compare model predictions to the data. 
All generated MC events are passed through the full
GEANT\cite{Brun:1987ma} based simulation of the H1 detector and are
reconstructed using the same analysis chain as is used for the data.

The {\sc HERWIG} 6.4 \cite{Marchesini:1991ch} MC event generator is used to
simulate the diffractive high $\tmod$ photon scattering using the LLA BFKL
model \cite{Ivanov:1998jw,Evanson:1999zb,Cox:1999kv}.
At leading logarithmic accuracy there are two independent free
parameters in the BFKL calculation: the value of the strong coupling 
$\alpha_s$ and the scale, $c$,
which defines the leading logarithms in the
expansion of the BFKL amplitude, $\ln(W^2/(c \ \tmod))$.
In exclusive production of vector mesons, the scale parameter $c$ is
related to the vector meson mass. 
In the case of diffractive photon scattering, the unknown scale 
results in the absence of a prediction for the normalisation of the
cross section~\cite{Cox:1999dw}.
In the calculations considered here \cite{Cox:1999kv},
the running of $\alpha_s$ as a function of the scale is ignored.
In order to distinguish the $\alpha_s$ parameter of the LLA
BFKL model from the usual strong coupling constant this parameter
will henceforth be referred to as $\alpeff$.
The choice of $\alpeff=0.17$ is used for the simulation in this Letter.

In the asymptotic approximation of the calculations
\cite{Cox:1999kv}, the W distribution follows a power-law:
\begin{equation}
 \nonumber
 \sigma(W) \sim  W^{4\omega_0},
 \label{eq:omega}
\end{equation}
where the exponent, also called the Pomeron intercept, is
given directly by the choice of $\alpeff$
with $\omega_0 = (3\alpeff/{\pi})\ 4\ \ln 2$. 
Using the {\sc HERWIG} simulation this approximation is
found to be justified given the current experimental statistical
precision.
The LLA BFKL model predicts an approximate power-law behaviour for the
$t$ dependence of the cross section of the form 
${\rm d}\sigma/{\rm d}\tmod \sim \tmod^{-n}$, where
$n$, also predicted by the model, depends only on the parton density
functions (PDFs) of the proton and the value of $\alpeff$. 
Running coupling effects, not considered here, would in principle 
allow $n$ to depend on $t$.

The GRV94 PDFs \cite{Gluck:1994uf} are used for the proton PDFs in
the {\sc HERWIG} prediction. A comparison with the CTEQ5 \cite{Lai:1999wy}
and MRST PDFs \cite{Martin:2002dr} shows
little dependence of the {\sc HERWIG} prediction on the input proton
PDFs. 

In order to describe the data, the $t$ dependence of the 
diffractive photon scattering simulated using {\sc HERWIG} 
(predicting a value of $n = 3.31$ for $\alpeff = 0.17$)
was weighted by a factor $|t|^{0.73}$, i.e.\
the $\tmod$ power is modified from $-3.31$ to $-2.58$.
This weighted {\sc HERWIG} prediction is used to correct the data for
resolution and acceptance effects.

Possible sources of background are estimated using MC simulations. 
The background from inclusive diffractive photoproduction events 
($e p \to e X Y$, where the two hadronic final states are separated by a
rapidity gap) is simulated using the {\sc PHOJET} MC event generator
\cite{Bopp:1998rc}. This background contributes when a single
electromagnetic particle fakes the photon candidate in the SpaCal.
It is estimated to amount to $3\%$ of the measured cross section.
The background from electron pair production ($ep \rightarrow ee^+e^-X$) is
modelled using the {\sc GRAPE} event generator \cite{Abe:2000cv}. 
This process contributes to the selection if one lepton is detected in the ET, 
a second lepton fakes the photon within the SpaCal and the remaining lepton 
escapes detection. This background contributes $4\%$ of the measured cross
section.

In order to investigate the background from
high $\tmod$ diffractive exclusive $\omega$ production, where the $\omega$
decays through the $\pi^+\pi^-\pi^0$ or $\pi^0\gamma$ channel, a
sample was generated using the {\sc DIFFVM} Monte Carlo generator
\cite{List:1998jz}.  The contribution from this background process is
found to be negligible. 
The background from DIS events, in which the scattered
positron fakes the photon candidate and an overlapping photoproduction
or Bethe-Heitler event gives a positron detected in the ET, has been studied 
and was also found to be negligible.  

\begin{figure}[!p]
\centering 
 \begin{picture}(200,170)(7,0)
  \put( 5,120){\epsfig{file=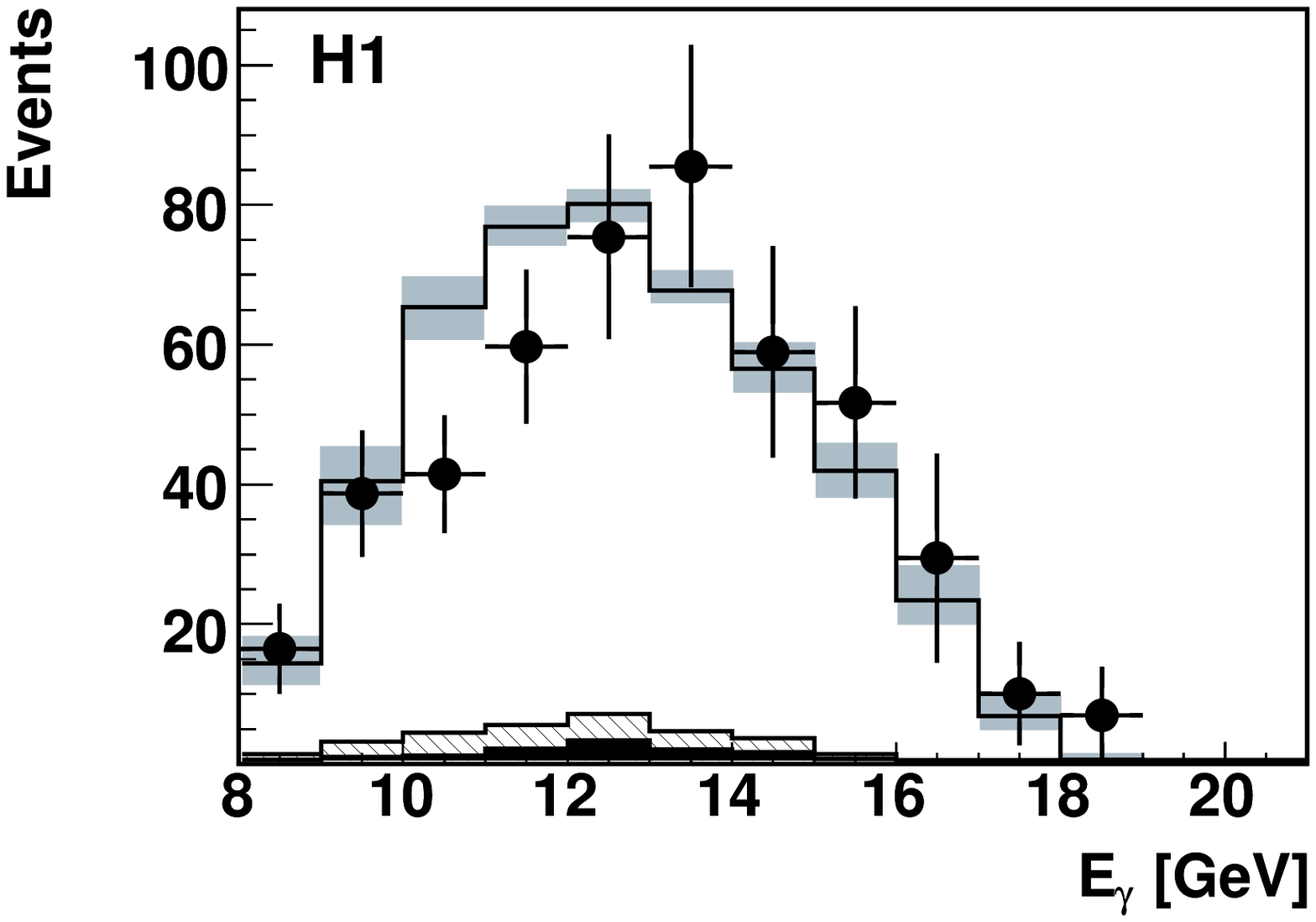,width=0.52\textwidth}}
  \put(91,120){\epsfig{file=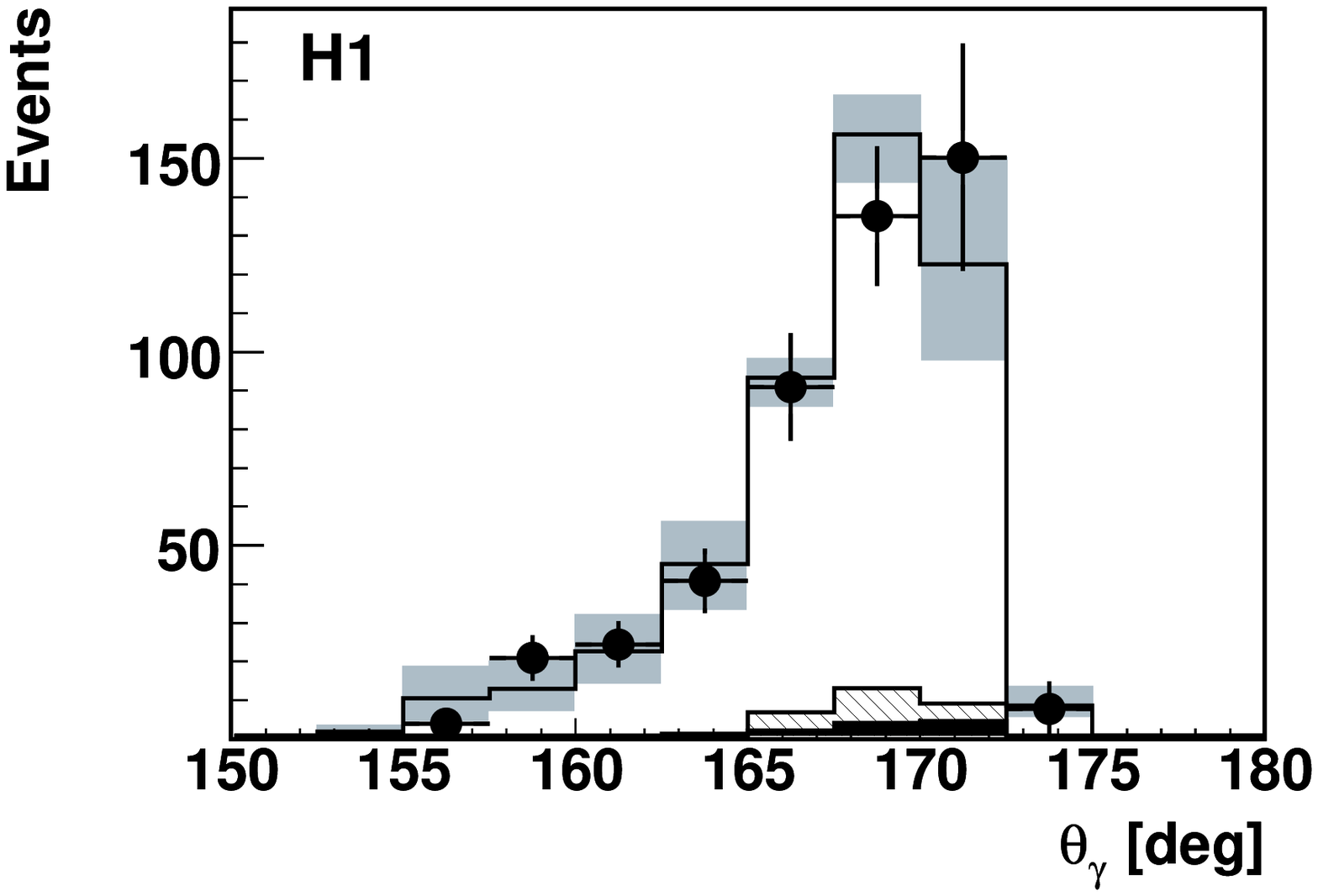,width=0.52\textwidth}}
  \put( 5, 58){\epsfig{file=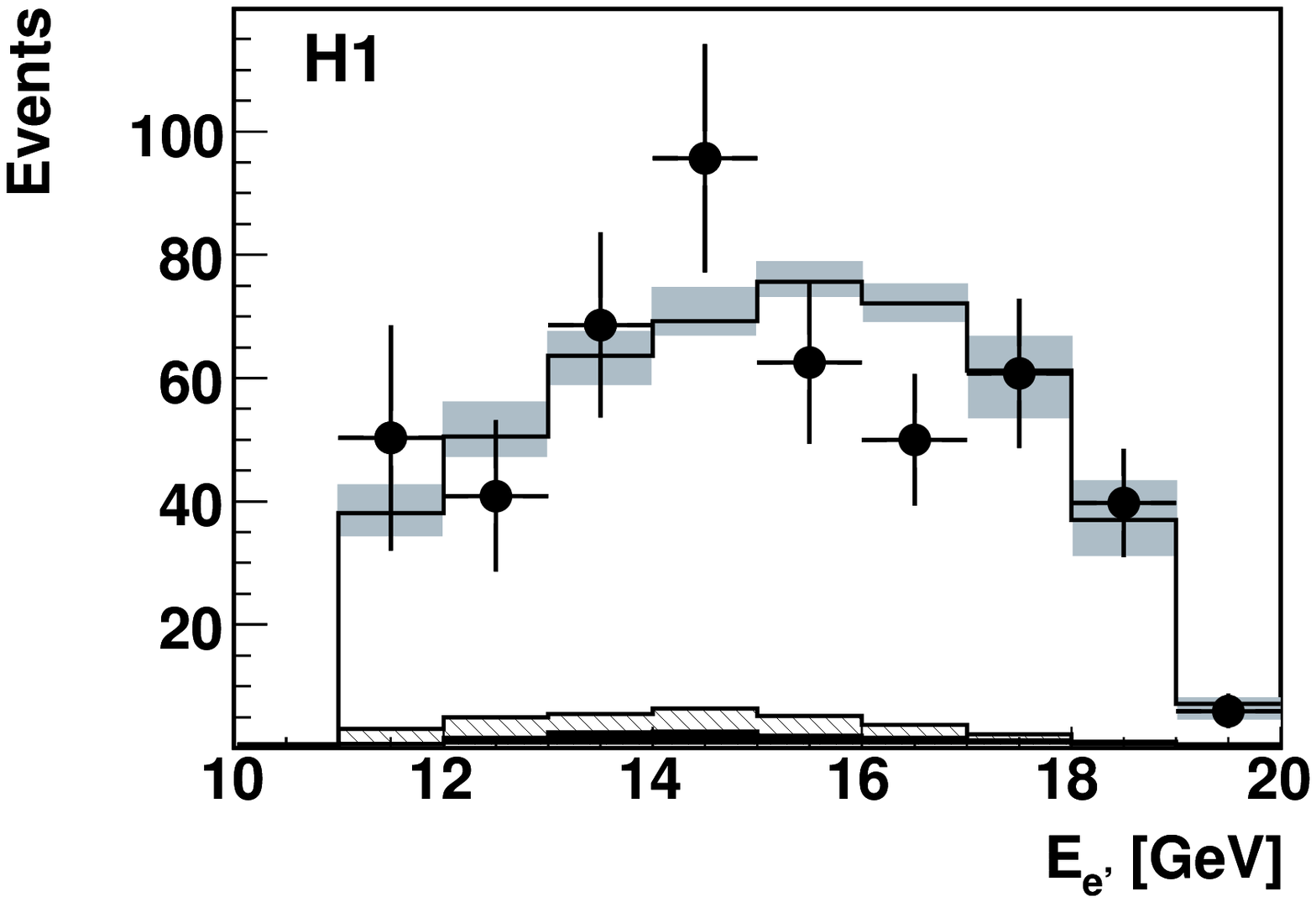,width=0.52\textwidth}}
  \put(91, 58){\epsfig{file=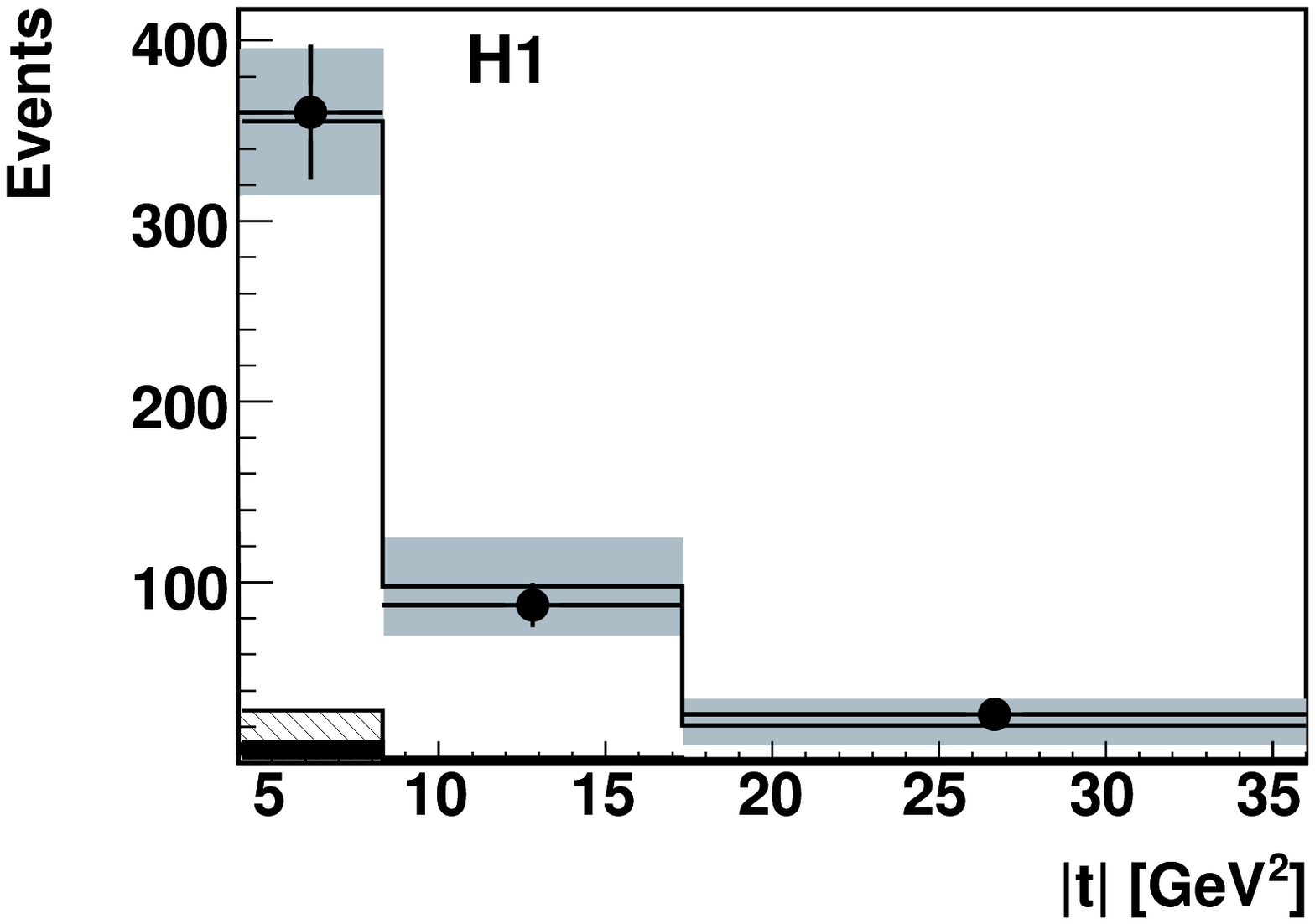,width=0.52\textwidth}}
  \put( 5, -4){\epsfig{file=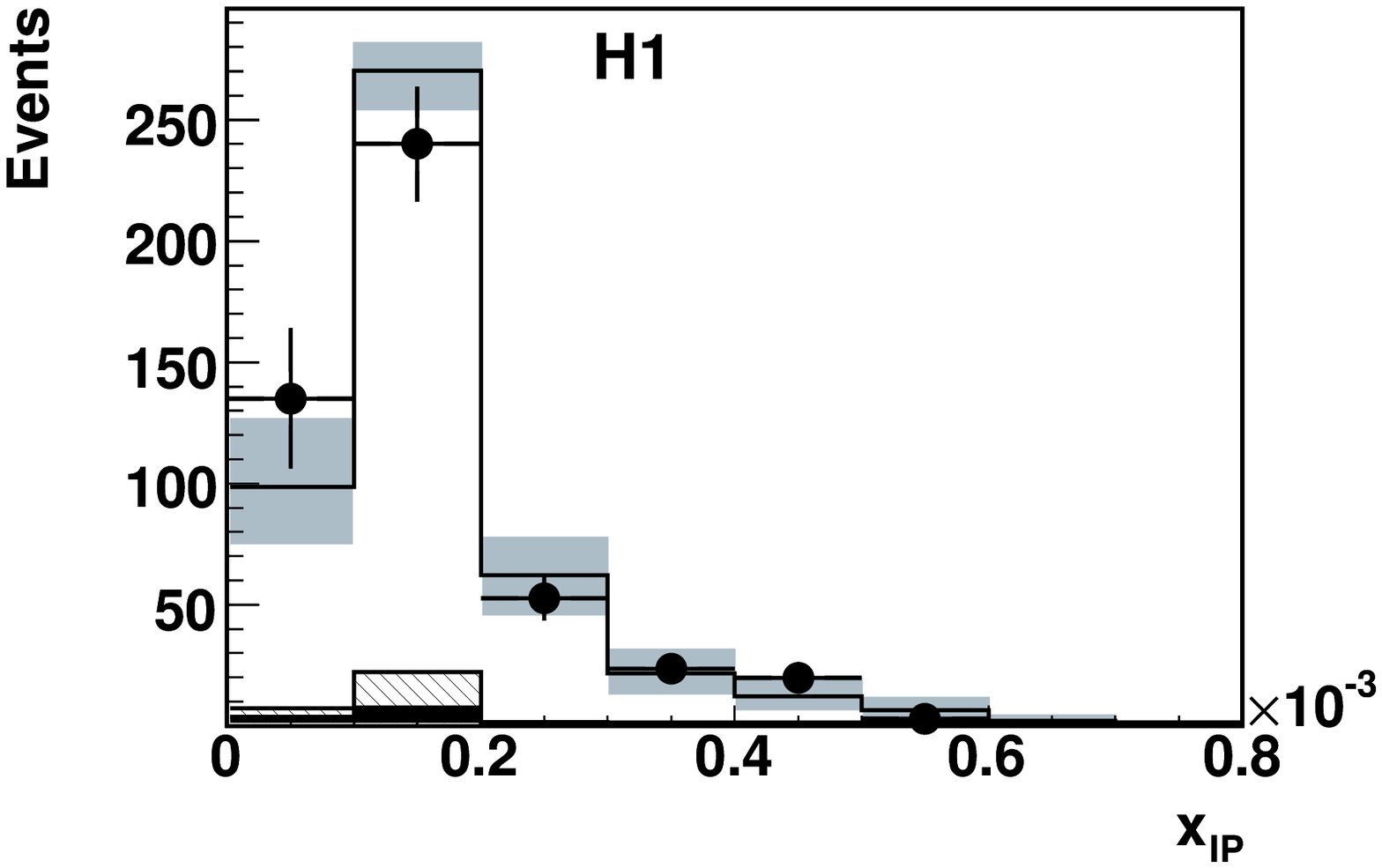,width=0.52\textwidth}}
  \put(91, -4){\epsfig{file=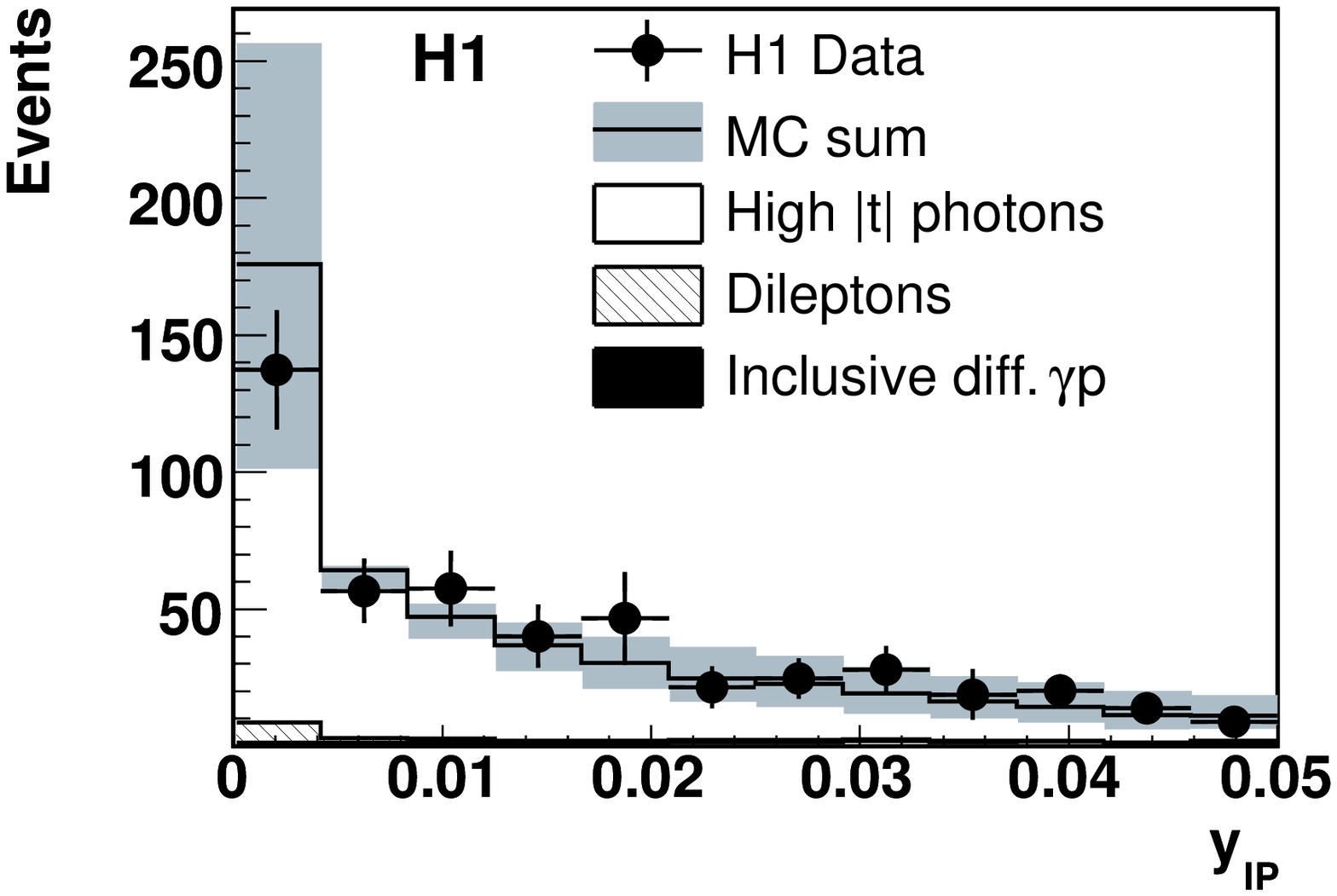,width=0.52\textwidth}}
  \put( 75,168){a)}
  \put(161,168){b)}
  \put( 75,106){c)}
  \put(161,106){d)}
  \put( 75, 44){e)}
  \put(161, 44){f)}
 \end{picture}

\caption{Distributions of the selected events as a function of
 a)~the final state photon energy, 
 b)~the final state photon polar angle, 
 c)~the scattered positron energy, 
 d)~\tmod, 
 e)~\xpom, and
 f)~\ypom.
 The data corrected for trigger efficiency (black points) are compared 
 with the simulation of diffractive high $\tmod$ photons from the  
 weighted {\sc HERWIG} based on the LLA BFKL calculation (open histogram), 
 and the background from inclusive
 diffractive photoproduction simulated with {\sc PHOJET} (full histogram) 
 and dilepton production simulated with {\sc GRAPE} (hatched histogram).
 The {\sc HERWIG} prediction is normalised to the number of data 
 events after background subtraction. The total systematic
 uncertainty of the simulation is shown by the dark grey shaded band.
}
\label{fig:control}
\end{figure}

In figure~\ref{fig:control} the data, corrected for trigger efficiency,
are compared to the Monte Carlo simulations.
The background predictions from {\sc GRAPE} and 
{\sc PHOJET} are normalised to the integrated luminosity
of the data sample. The weighted {\sc HERWIG} prediction is normalised to 
the number
of events obtained after the subtraction of the predicted background from the data. 
A good
description of the data by the combined Monte Carlo simulations is observed.

\section{Cross Section Measurement}

The $e p \to e \gamma Y$ differential cross sections are defined using 
the formula:
\begin{equation}
 \nonumber
 \frac{{\rm d^2}\sigma_{ep \to e \gamma Y}}{{\rm d} W \ {\rm d}t} 
  = \frac{N_{data} - N_{bgr}}{{\cal L} A\, \Delta W \, \Delta t},
\end{equation}
where $N_{data}$ is the number of observed events corrected for trigger
efficiency, $N_{bgr}$ the 
expected contribution from background events as estimated using the
{\sc PHOJET} and {\sc GRAPE} Monte Carlos simulations, ${\cal L}$ the 
integrated luminosity, $A$ the signal acceptance and $\Delta W$ and 
$\Delta t$ the bin widths in $W$ and $t$, respectively. 
The acceptance, estimated using the weighted {\sc HERWIG} simulation,
is the ratio of the number of events accepted after
reconstruction to the number of events generated in
the defined phase space on hadron level. 
It accounts for all detector effects, including bin-to-bin migrations, as
well as geometrical acceptance and detector efficiencies.
QED radiative correction effects are estimated to be less than $1$\% 
\cite{DESY-95-162} and are neglected.

The $\gamma p \to \gamma Y$ differential cross
section is then extracted from the $ep$ cross section using:
\begin{equation}
 \frac{{\rm d^2}\sigma_{ep \to e \gamma Y}}{{\rm d} W \ {\rm d}t} 
  = \Gamma(W) \ \frac{{\rm d}\sigma_{\gamma p \to \gamma Y}}
    {{\rm d}t}(W),
 \label{eq:flux}
\end{equation}
where the photon flux, $\Gamma(W)$, is integrated over the range
$Q^2<0.01$ GeV$^2$ according to the modified Weizs\"acker-Williams 
approximation \cite{vonWeizsacker:1934sx}.
The $\gamma p$ cross section is obtained by modelling 
$\sigma_{\gamma p \to \gamma Y}$ as a power-law in $W$, 
whose parameters are iteratively adjusted to reproduce the measured 
$W$ dependence of the $ep$ cross section. 
The differential $\gamma p$ cross section in $\tmod$ is then extracted 
from the $ep$ cross section by correcting for the effect of the photon flux
over the visible $W$ range ($175 < W < 247$ GeV).
The $\gamma p$ cross section as a function of $W$ is obtained by 
first integrating equation~(\ref{eq:flux}) over the $\tmod$ range, 
and then correcting for the effect of the photon flux in each bin in $W$.
More details on the extraction procedure of the $\gamma p$ cross section 
can be found in \cite{Hreus08}.

The systematic error on the measurement stems from experimental
uncertainties and from model dependences.
They are calculated using the weighted {\sc HERWIG}
simulation of the signal process. 
The sources of systematic error are listed below. For each of them the
typical effect on the cross section measurement is indicated. 

\noindent The experimental systematic errors are:
\begin{itemize}
 \item The energy scale uncertainty of $\pm 1 \%$ for an electromagnetic 
       cluster measured by the SpaCal gives errors in the range of $2\%$ 
       to $4\%$.
 \item The uncertainty of $\pm2.5$~mrad for the measurement of the photon
       candidate polar angle results in an error of up to $3\%$.
 \item The hadronic final state energy scale uncertainty of $\pm 4 \%$  
       leads to an error of less than $1.5\%$.
 \item The energy scale uncertainty of $\pm 1.5 \%$ of the ET produces
       an error ranging from $1\%$ in the highest $\tmod$ bin to $10\%$ in 
       the lowest $W$ bin.
 \item The uncertainty of $\pm 25\%$ on the noise threshold from the 
       calorimeters gives an error varying from $5\%$ at low $\tmod$ to $11\%$
       at the highest $\tmod$ value.
 \item The luminosity is measured with an accuracy of $\pm 1.5 \%$
       which enters as an overall normalisation uncertainty.
\end{itemize}
\noindent The systematic errors due to uncertainty of model parameters
are:
\begin{itemize}
 \item The uncertainty due to the \xpom dependence, estimated by weighting
       the \xpom distribution by the form $(1/\xpom)^{\pm 0.4}$, yields 
       an error varying from $3\%$ in the central $\tmod$ bin up to $9\%$ in
       the lowest $W$ bin. 
       This weight would correspond to changing $\alpeff$ to the
       values $0.02$ and $0.31$.
 \item The uncertainty due to the $\tmod$ dependence, estimated by weighting
       the $\tmod$ distribution by the form $(1/\tmod)^{\pm 0.2}$, leads to 
       an error ranging from $1.5\%$ to $4\%$.
 \item The uncertainty in the modelling of the proton remnant system~$Y$, 
       estimated by weighting the $M_Y$ distribution by the form 
       $(1/M_Y^2)^{\pm 0.3}$, results in a typical error of $1\%$ to $4\%$.
 \item The uncertainty of $100\%$ assumed on the normalisation of the 
       subtracted inclusive diffractive background (from the {\sc PHOJET} Monte
       Carlo simulation) leads to an error of approximately $3\%$ in the highest
       $W$ bins decreasing to $1\%$ in the highest $\tmod$ bin.
 \item The propagation of the uncertainty on the power-law parameter $\delta$ in the 
       $\gamma p$ cross section extraction procedure leads to an error
       of $4\%$ on ${\rm d}\sigma_{\gamma p \to \gamma Y}/{\rm d}\tmod$,
       independent of $\tmod$, 
       and is below 1\% on $\sigma_{\gamma p \to \gamma Y}$.
\end{itemize}

The uncertainty on the PHOJET MC normalisation and the model uncertainties from
the unknown \xpom, $\tmod$ and $M_Y$ dependences are estimated from 
data comparisons and defined by the range where the model
describes the data.
%
%
Each source of systematic error is varied in the weighted HERWIG Monte
Carlo within its uncertainty. In each measurement bin, the corresponding
deviation of the normalised cross section from the central value is
taken as systematic error.
The total systematic error is obtained by adding the individual
contributions in quadrature, on a bin-by-bin basis.
The largest systematic error on the cross sections comes 
from the uncertainty on the \xpom and $M_Y$ dependences in the MC
simulation.  
The total systematic error on the $W$ dependence of the cross section varies
from $10\%$ in the central bins to $17\%$ in the lowest bin. 
The systematic error on the $\tmod$ dependence of the cross section varies 
from $8\%$ in the lowest bin to $14\%$ in the highest bin. An additional
global uncertainty of 4\% arises from the $\gamma p$ cross section
extraction procedure. 
The total systematic errors are comparable to or smaller than the 
statistical errors. 
 
\section{Results} 
 
The cross sections measured for the domain $175 < W < 247~\mathrm{GeV}$,
$4 < \tmod < 36$ GeV$^2$, $\ypom < 0.05$ and $\qsq<0.01\gevsq$
are presented in table \ref{CSvalues}.

\begin{figure}[!p]
 \centering 
 \epsfig{file=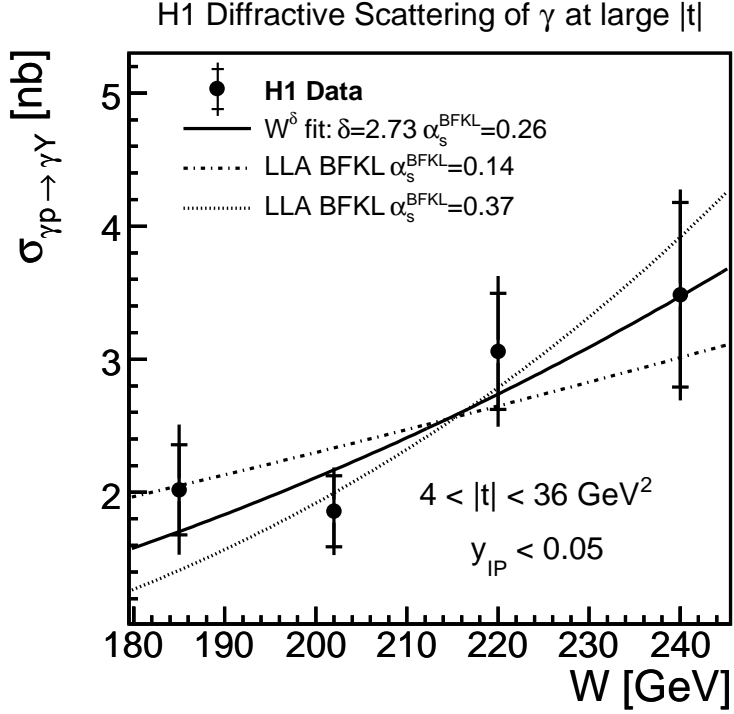,width=0.68\textwidth}
 \caption{The $\gamma p$ cross section of diffractive scattering of
  photons as a function of $W$ in the phase space defined by 
  $4 < \tmod < 36$\gevsq , $\ypom < 0.05$ and $\qsq<0.01\gevsq$.
  The average value is $\langle \tmod \rangle =6.1$ GeV$^2$.
  The inner error bars show the statistical
  errors and the outer error bars show the statistical and systematic
  errors added in quadrature. The solid line shows the result of 
  a fit to the cross section of the form $W^\delta$ with $\wfitres$.
  This line also corresponds to the LLA BFKL model prediction from the 
  {\sc HERWIG} event generator with $\alpeff = 0.26$.
  Two additional curves show the LLA BFKL predictions for the additional
  choices of $\alpeff = 0.14$ and $0.37$ corresponding to one standard 
  deviation from the fit value. 
 }
 \label{fig:SigmaW}
\end{figure}

The $\gamma p \rightarrow \gamma Y$ cross section as a function 
of $W$ is shown in figure~\ref{fig:SigmaW}. 
A power-law dependence of the form
$\sigma \sim W^{\delta}$ is fitted to the measured cross section. 
The fit yields $\wfitres$ with $\chi^2/\textrm{n.d.f.}=2.7/2$. The
contributions from the systematic errors are calculated by
shifting the data points according to each source of uncertainty, 
taking correlations into account, and
repeating the fit. The errors are then added in quadrature
to obtain the total systematic error. 

The steep rise of the cross
section with $W$ is usually interpreted as an indication of the
presence of a hard sub-process in the diffractive interaction and of
the applicability of perturbative QCD. 
The present $\delta$ value, measured at an average $\tmod$
value of $6.1$~GeV$^2$,
is compatible with that measured by H1 in diffractive
$J/\psi$ photoproduction of $\delta=1.29 \pm0.23 (\mathrm{stat.})\pm
0.16(\mathrm{syst.})$ \cite{Aktas:2003zi} at an average $\tmod$ of
$6.9$~GeV$^2$. 
The Pomeron intercepts associated with these $\delta$ values 
correspond to the strongest energy dependences 
measured in diffractive processes.

\begin{table}
\begin{center}
\begin{tabular}{|c|c|c|c|c|}
\hline
\multicolumn{5}{|c|}{H1 measured $ep \rightarrow e \gamma Y$ cross sections}\\
\hline
$W$ & $W$ range & ${\rm d}\sigma_{ep \rightarrow e \gamma Y}/{\rm d}W$ &
$\Gamma(W)$ & $\sigma_{\gamma p \rightarrow \gamma Y}$ \\
~[GeV] & [GeV] & $[$pb/GeV$]$ & [GeV$^{-1}$] & [nb] \\
\hline
$185$ & $175-193$ & $0.414 \pm 0.069 \pm 0.072$ & $0.0565$ & $2.02  \pm 0.34  \pm 0.35$\\
$202$ & $193-211$ & $0.318 \pm 0.046 \pm 0.033$ & $0.0431$ & $1.86  \pm 0.27  \pm 0.19$\\
$220$ & $211-229$ & $0.434 \pm 0.062 \pm 0.051$ & $0.0328$ & $3.06  \pm 0.44  \pm 0.36$\\
$240$ & $229-247$ & $0.404 \pm 0.080 \pm 0.044$ & $0.0246$ & $3.48  \pm 0.69  \pm 0.38$\\
\hline
\hline
$|t|$ & $|t|$ range & d$\sigma_{ep \rightarrow e \gamma Y}/{\rm
d}|t|$ & $\Gamma(W=219~\mathrm{GeV})$ & d$\sigma_{\gamma p \rightarrow \gamma Y}/$d$|t|$\\
~[GeV$^2$]& [GeV$^2$]& $[$pb/GeV$^2]$ & [GeV$^{-1}$] & [pb/GeV$^2$] \\
\hline
$ 6$ & $4.0-8.3 $  & $4.04 \pm 0.42 \pm 0.36$ & $0.0333$ & $401 \pm 41 \pm 36$ \\
$12$ & $8.3-17.3$  & $0.58 \pm 0.08 \pm 0.06$ & $0.0333$   & $ 57.8 \pm 8.1 \pm 6.2 $\\
$25$ & $17.3-36.0$ & $0.13 \pm 0.03 \pm 0.02$ & $0.0333$   & $ 12.5 \pm 3.1 \pm 1.8 $\\
\hline
\end{tabular}
\caption{The cross sections for the processes
$ep \rightarrow e \gamma Y$ and the $\gamma p \rightarrow \gamma Y$, 
measured in the range $\ypom < 0.05$ and $\qsq<0.01\gevsq$. 
The upper part of the table presents the measured cross
sections for different values of $W$ at an average $\langle \tmod \rangle =6.1$ GeV$^2$.
The lower table presents the measured cross sections differential in $\tmod$ 
at $W=219$ GeV.
The first errors are statistical, the second systematic.
The photon flux $\Gamma$ and corresponding ranges in $W$ and $\tmod$ 
used for the measurements are also given.}
\label{CSvalues}
\end{center}
\end{table}

\begin{figure}[!p]
 \centering 
 \epsfig{file=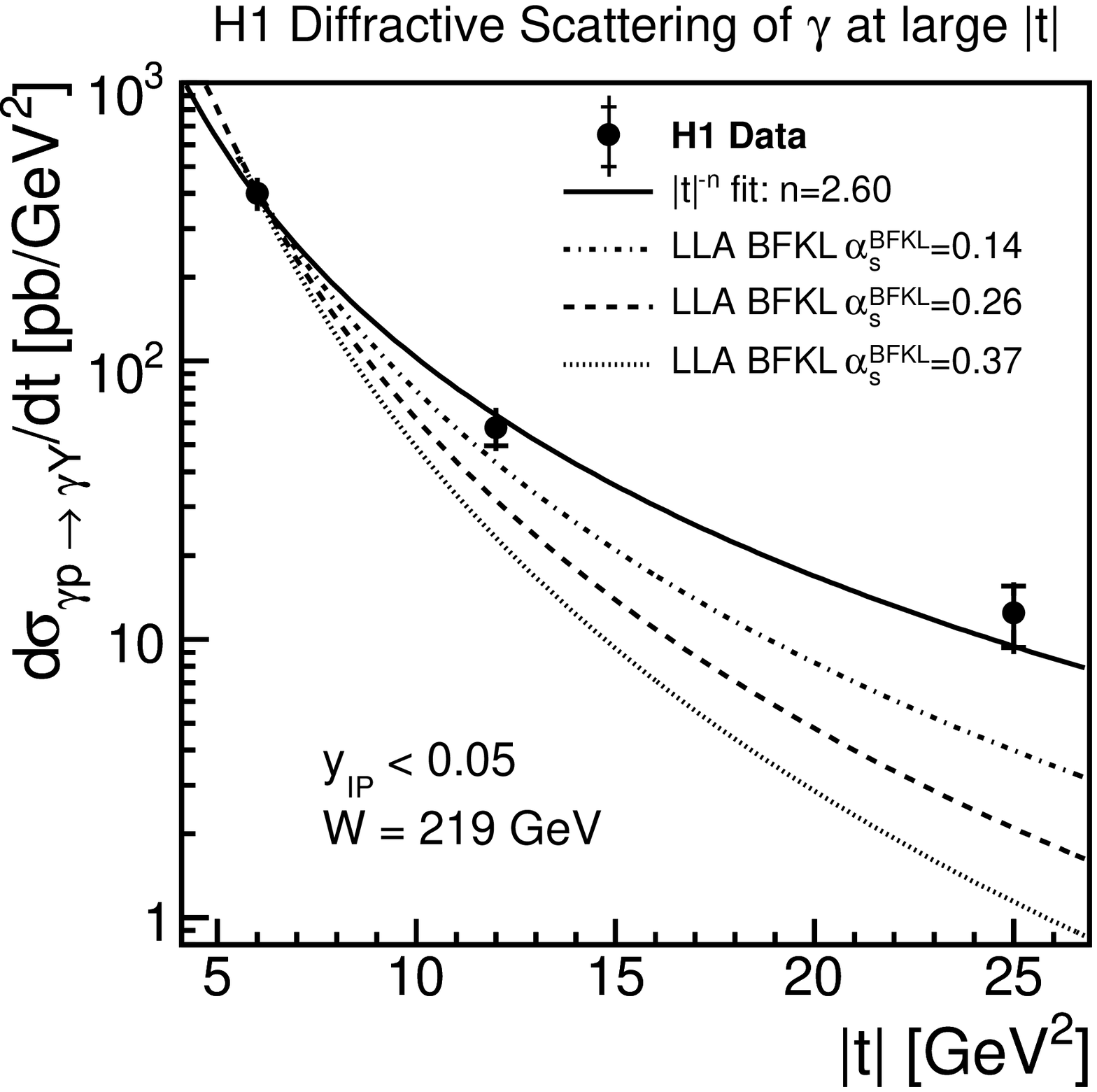,width=0.68\textwidth}
 \caption{The $\gamma p$ cross section of diffractive scattering of
  photons differential in $\tmod$ for $W=219$~GeV, 
  $\ypom < 0.05$ and $\qsq<0.01\gevsq$.
  The inner error bars show the statistical
  errors and the outer error bars show the statistical and systematic
  errors added in quadrature. The solid line shows the result of 
  a fit to the cross sections of the form $\tmod^{-n}$ with $\tfitres$.   
  Three additional curves show the LLA BFKL model predictions from the
  {\sc HERWIG}
  event generator for the values $\alpeff = 0.14, 0.26$ and $0.37$.
 }
 \label{fig:Sigmat}
\end{figure}

The $\gamma p$ cross section differential in $\tmod$, at $W=219$ GeV, 
is shown in figure~\ref{fig:Sigmat}. 
Figure \ref{fig:Sigmat} also shows the comparison of the cross 
section to a fit of the form ${\rm d}\sigma/{\rm d}t \sim \left|t\right|^{-n}$. 
The fit result is $\tfitres$ with $\chi^2/\textrm{n.d.f.}=1.6/1$. 
The $\tmod$ dependence is harder than that 
measured by H1 in the diffractive photoproduction of $J/\psi$ mesons at large $|t|$
\cite{Aktas:2003zi}
corresponding to $n=3.78\pm0.17(\mathrm{stat.})\pm 0.06(\mathrm{syst.})$.

In figures~\ref{fig:SigmaW} and \ref{fig:Sigmat} the measured cross 
sections are compared to predictions of the LLA BFKL model, using the
{\sc HERWIG} Monte Carlo, as described in section \ref{sec:mc}, 
with no $\tmod$ weighting. 
The predictions are normalised to the integrated measured cross
section, as the normalisation is not predicted by the LLA BFKL
calculation \cite{Cox:1999dw}. 
The $W$ dependence of the cross section is well described by the LLA 
BFKL prediction, as shown in figure~\ref{fig:SigmaW}.
The sensitivity of the measurement to the free parameter
$\alpeff$ in the theoretical prediction is illustrated in
figures~\ref{fig:SigmaW} and \ref{fig:Sigmat}.
Using $\delta= 4\, \omega_0 = 4\ (3\alpha_S^{Fit}/{\pi})\ 4\ \ln 2$, 
the measured $W$ dependence leads to $\alphasfitres$.
Predictions are shown in figure~\ref{fig:SigmaW} for 
the values $\alpeff=0.14$ and $0.37$. 
The LLA BFKL curve corresponding to $\alpeff=0.26$ coincides with 
the solid fit line.

Previous measurements of diffractive scattering at HERA 
are well described by BFKL predictions with
$\alpeff$ values similar to the value measured in this Letter.
ZEUS measurements of exclusive $\rho, \phi$ and $J/\psi$ production
at large $|t|$ are best described with the value of
$\alpeff=0.20$ \cite{Chekanov:2002rm}.  The H1
measurement of high $\tmod$ $\rho$ production \cite{Aktas:2006qs} is
compatible with $\alpeff=0.20$ and
the H1 measurement of high $\tmod$ $J/\psi$ production
\cite{Aktas:2003zi} is described using $\alpeff=0.18$.
The LLA BFKL prediction with $\alpeff =0.17$ gives a
good description of the double dissociation
process with a rapidity gap between jets measured by H1
\cite{Adloff:2002em}. Events with rapidity gaps between jets were also measured
by ZEUS \cite{Chekanov:2006pw} and are found to be compatible
with a model which uses $\alpeff =0.11$. Note, however, that for these 
measurements the hard scale, corresponding to the jet transverse momentum
squared, is of the order of or larger than $20$ GeV$^2$.

As shown in figure~\ref{fig:Sigmat}, the LLA BFKL calculation
for $\alpeff=0.14, 0.26$ and $0.37$, all of which give
a reasonable description of the $W$ dependence, predict steeper $\tmod$
distributions than is measured in the data.
The same effect cannot be established for the exclusive $\rho$
measurement \cite{Aktas:2006qs}, where the measured $t$ range is limited to 
$\tmod < 8$ GeV$^2$, although an underestimate of the cross section was
observed at the largest values of $\tmod$. 
The present situation is in contrast with the analysis of $J/\psi$
production \cite{Chekanov:2002rm,Aktas:2003zi}, where the $\tmod$ dependence 
was found to be well described by the LLA BFKL prediction over a similar
range in $t$.


\section{Conclusions}

Using the H1 detector, 
diffractive photon scattering, $\gamma p\rightarrow \gamma Y$, where
the final state photon carries a large transverse momentum and is well
separated from the proton dissociative system~$Y$, is measured
for the first time at HERA. The measurement provides
a unique test of the underlying QCD dynamics of the
diffractive exchange.

Cross sections are presented as a function of $W$ 
and differentially in $|t|$. A fit of the form $W^\delta$ performed on the
cross section gives \wfitres. This strong energy dependence is compatible 
with that measured for the exclusive diffractive $J/\psi$ production
at high $\tmod$.
A fit of the form $\tmod^{-n}$ yields \tfitres, corresponding to a 
harder $\tmod$ dependence of the cross section than measured
for high $\tmod$ diffractive $J/\psi$ production.

The measured cross sections are compared to the predictions of an LLA BFKL
model.
A good description of the $W$ dependence of the cross section is found,
whereas the LLA BFKL model predicts a $\tmod$ dependence that is too soft 
and hence unable to describe the data.

\section*{Acknowledgements}

We are grateful to the HERA machine group whose outstanding efforts
have made this experiment possible.  We thank the engineers and
technicians for their work in constructing and maintaining the H1
detector, our funding agencies for financial support, the DESY
technical staff for continual assistance and the DESY directorate for
support and for the hospitality which they extend to the non DESY
members of the collaboration.
We are also grateful to J.R.\ Forshaw and M.\ Diehl for fruitful
discussions.


\end{document}